\documentclass{article}
\usepackage{amssymb}
\usepackage{amsmath}
\usepackage{amsfonts}
\usepackage{hyperref}
\usepackage{multimedia}
\usepackage{graphicx}
\usepackage{setspace}
\usepackage{subfigure}
\usepackage{rotating}

\newtheorem{theorem}{Theorem}
\newtheorem{acknowledgement}[theorem]{Acknowledgement}

\input{tcilatex}
\begin{document}

\title{Blockchains and Distributed Ledgers in Retrospective and Perspective}
\author{Alexander Lipton \\
Stronghold Bank Labs\\
MIT Connection Science and Engineering}
\maketitle

\begin{abstract}
We introduce blockchains and distributed ledgers and describe their
potential applications to money and banking. The analysis compares public
and private ledgers and outlines the suitability of various types of ledgers
for different purposes. Furthermore, a few historical prototypes of
blockchains and distributed ledgers are presented, and results of their hard
forking are illustrated. Next, some potential applications of distributed
ledgers to trading, clearing and settlement, payments, trade finance, etc.
are outlined. Monetary circuits are argued to be natural applications for
blockchains. Finally, the role of digital currencies in modern society is
articulated and various forms of digital cash, such as central bank issued
electronic cash, bank money, bitcoin and P2P money, are compared and
contrasted.

\textit{Keywords}: blockchains, distributed ledgers, digital currencies,
modern monetary circuit; credit creation banking; interconnected banking
network.
\end{abstract}

\tableofcontents

\begin{quote}
"PARATOV. The madness of passion soon passes, and what remains are chains
and common sense that tells us that these chains are unbreakable. LARISA.
Unbreakable chains!"

Alexander Ostrovsky, Without a Dowry, A drama in four acts
\end{quote}

\section{Introduction}

\label{Introduction}In this paper, we discuss blockchains (BCs) and
distributed ledgers (DLs) in retrospective and prospective, with an emphasis
on their applications to money and banking in the 21st century. Additional
aspects are discussed in \cite{Lipton1}, \cite{LiptonShrierPentland}, \cite%
{Tasca}.

Civilization is not possible without money, and, by extension, banking, and
vice versa. Through the ages, money existed in many forms, stretching from
the exquisite gold coins of the Phrygian King Midas, giant stones of
Polynesia, cowry shells, the paper money of Khublai Khan and other rulers
who came after him, to digital currencies, and everything in between. The
meaning of money has preoccupied rulers and their tax collectors, traders,
entrepreneurs, laborers, economists, philosophers, writers, stand-up
comedians, and ordinary folks alike. It is universally accepted that money
has several important functions, such as a store of value, a means of
payments in general, and taxes in particular, and a unit of account. The
author shares the view of Aristotle formulated in his \textit{Ethics}:
"Money exists not by nature but by law." Thus, money is linked to government
and government to money. In fact, anything taken in lieu of tax eventually
becomes money.

For the last five centuries, money has gradually assumed the form of records
in various ledgers. This aspect of money is all-important in the modern
world.\ At present, money is nothing more than a sequence of transactions,
organized in ledgers maintained by various private banks, and by central
banks who provide means (central bank cash) and tools (various money
transfer systems) used to reconcile these ledgers. In addition to their
ledger-maintaining functions, private banks play a very important role,
which central banks are not equipped to perform. They are the system
gatekeepers, who provide know your customer (KYC) services, and system
policemen, who provide anti-money laundering services (AML). We argue that,
in addition to the more obvious areas of application of distributed ledger
technology (DLT), for instance, digital currencies (DC), including central
bank issued digital currencies (CBDC), DLT can be used to solve such complex
issues as trust and identity, with an emphasis on the KYC and AML aspects, 
\cite{Zyskind}. Further, given that all banking activities boil down to
maintaining a ledger, judicious applications of DLT can facilitate trading,
clearing and settlement triad, payments, trade finance, etc.

The paper is organized as follows. We introduce DLs and briefly discuss
their different types in Section \ref{DLTgeneral}. We present historical
instances of BCs and DLs in Section \ref{Examples}, and describe what
happened when they underwent hard forking. Bitcoin, the most popular current
application of DLT, is covered in Section \ref{Bitcoin}, where a\ few less
well known facts about bitcoin are presented. Potential applications of DLT
to banking are discussed in Section \ref{DLTinbanking}. As an interesting
potential area of applications of BC/DLT, we introduce modern version of
monetary circuit in Section \ref{Circuit} and show that it can benefit from
the BC/DL framework because money moves in a gigantic circle (or several
circles if the world economy as a whole is considered). In addition, in the
process of money creation by the banking system as a whole, individual banks
become naturally interconnected, so that DLs are particularly suitable to
describing their interactions. We discuss topics related to CBDC in Section %
\ref{CBDC}, where we explain the rational for its issuance and discuss
practical aspects. In particular, we show that CBDC can be used to implement
the famous Chicago plan, \cite{Allen}, \cite{Benes}, of moving away from the
fractional reserve banking toward the narrow banking. We articulate the
differences between Chaum's and Nakamoto's approaches to DC and consider
their respective pros and cons. Conclusions are drawn in Section \ref%
{Conclusions}.

\section{BCs and DLs}

\label{DLTgeneral}Databases with joint writing access have been known for
decades. Several typical examples are worth mentioning: the concurrent
versioning system (CVS), Wikipedia, and distributed databases used on board
of naval ships \cite{Miller}.

We start with articulating differences between centralized and distributed
databases. In a centralized database, storage devices are all connected to a
common processor; in a distributed database, they are independent.
Furthermore, in a centralized database, writing access is tightly
controlled; in a distributed database, many actors have writing privileges.
In the latter case, each storage device maintains its own growing list of
ordered records, which, for the efficiency sake, can be organized in blocks,
hence, the name Blockchain. To put it differently, in a traditional
centralized ledger, the gatekeeper collects, verifies, and performs the
write requests of multiple parties, tasks which are distributed in the DL.
It should not be taken as fact that these tasks are best distributed: the
considerations of efficiency and specialization are relevant as well.

The integrity of the distributed database is cryptographically ensured at
two levels. First, only users possessing private keys, can make updates to
"their" part of the ledger. Second, notaries (also called miners) verify
that users' updates are legitimate. Once the updates are notarized, they are
broadcast to the whole network, thus ensuring that all copies of the
distributed database are in sync. There several types of distributed
databases or ledgers. We list them in increasing order of complexity:

(A) traditional centralized ledger;

(B) permissioned private DL (R3 CEV, DAH, and other similar projects);

(C) permissioned public DL (Ethereum, Ripple, etc.);

(D) unpermissioned public DL (bitcoin and the myriad others).

To control the integrity of DL, a variety of mechanisms can be used - proof
of work (PoW), proof of stake (PoS), third party verification, etc.

Which ledger should be used? It largely depends on the context. If no joint
writing access is required, as is the case with most legacy banking
applications, a centralized ledger can be used. If participants do need
joint writing access, but know each other in advance, have aligned
interests, and can be trusted, as is the case in clearing and settlement, a
permissioned private DL can be employed. More details are given in \cite%
{Greenspan}.

The best known application of BC/DL is the famous bitcoin, which exists on
an unpermissioned public DL whose integrity is maintained by anonymous
miners via PoW mechanism. BC/DL can be used for issuing CBDC. However, the
sheer scale of the economy precludes unpermissioned public ledger in the
spirit of S. Nakamoto, \cite{Nakamoto} to be used for this purpose, due to
the enormous computational effort required for PoW. Resurrecting digicash
proposed by D.\ Chaum, \cite{Chaum}, is an exciting possibility.

In many instances, building a DL just to be \textit{au courant} with times
might not be worth the effort.

\section{Historical examples of BCs and DLs}

\label{Examples}

\subsection{Genealogical trees}

The idea of a BC is certainly not new. BCs naturally occur whenever power,
land, or property change hands. Some of the earliest examples of BC are the
genealogical trees of royal (or, more generally, aristocratic or property
owning) families. In such a tree (or BC) the transfer of power from one
sovereign to the next is governed by well-defined rules and in most cases,
occurs without commotion. However, when these rules become ambiguous and
open to interpretation the tree can undergo a hard fork.

In addition to being a chain, a genealogical tree is a distributed ledger.
In order to agree on their respective legitimacy and marriage eligibility,
royal houses had to inform each other about births, deaths, marriages, and
other life events, thus keeping their versions of BCs in sync. In Figure \ref%
{Fig 1} we show the genealogical tree of the House of Habsburg engraved by
A. Durer. It was distributed to other royal houses, as well as all imperial
cities in the Holy Roman Empire.

Usually forking of a succession tree is associated with wars and other acts
of violence. This is a cautionary tale for proponents of ubiquitous
applications of DLs without a possibility of resolving disputes outside of
the ledger itself. Here are two (of many) examples.

In Figure \ref{Fig 2} we show a simplified genealogical tree of the House of
Capet. For ten generation, starting with Hugh Capet, the transfer of power
from father to son was smooth. However, the ambiguity occurred when all
three sons of Philip IV died without surviving issue, thus creating a power
vacuum. In order to resolve it, the peers of France applied the Salic law of
Succession, by which persons descended from a previous sovereign only
through a woman are not eligible to occupy the throne. The House of
Plantagenet did not accept this outcome and started the Hundred Years' War
(1337-1453) against the House of Valois, a cadet branch of the Capetian
dynasty, which was a dynastic conflict for control of the Kingdom of France.
In the end, the Valois, established themselves as Kings of France at the
expense of the Plantagenets.

Similar conflicts occurred with regularity and for very similar reasons
throughout the history. For example, the War of the Austrian Succession
(1740--1748), which involved all major powers of Europe, was fought to
settle the question of the Pragmatic Sanction and decide whether the
Habsburg hereditary possessions could be inherited by a woman. It was
finally resolved in favor of Maria Theresa, who became the only female ruler
of the Habsburg dominions.

Closer to our times, an interesting example of Ethereum hard forking
happened in July of 2016, as a result of fixing a theft of \$60 Mil worth of
Ethereum from DAO. V. Buterin, \cite{Buterin}, described the situation as
follows:

\begin{quote}
"The foundation has committed to support the community consensus on the
admittedly difficult hard fork decision. ... That said, we recognize that
the Ethereum code can be used to instantiate other blockchains with the same
consensus rules, including testnets, consortium and private chains, clones
and spinoffs, and have never been opposed to such instantiations."
\end{quote}

Once again, we see that ambiguity within a BC cannot be resolved via its
intrinsic mechanisms.

\subsection{Land titles}

In more recent times, land registry title deeds are more relevant examples
of blockchains. As per Land Registry,

\begin{quote}
"Title deeds are paper documents showing the chain of ownership for land and
property. They can include: conveyances, contracts for sale, wills,
mortgages and leases."
\end{quote}

It is clear that titles are blockchains currently held in a central
repository; however, instead of miners, succession is verified by notaries.
Titles are meaningful candidates for being treated on DL. However, there are
still some issues which need to be resolved before it can be done. For
example, recent lawsuits by Mark Zuckerberg seeking to force hundreds of
Hawaiians to sell to him small plots of land located within the external
boundaries of his 700-acre beachfront property on the island of Kauai, is a
good case in point. It illustrates that in some instances, it is not
possible to identify the first owner of land, and then build a chain of
ownership from the original owner to the present, resulting in an ambiguous
and potentially vulnerable BC.

\section{The bitcoin ecosystem}

\label{Bitcoin}Bitcoin is not the first digital currency by a long shot, and
very likely is not the last major one either. The astute reader will
recognize that apart from intriguing technical innovations, bitcoin does not
differ that much from the fabled tally sticks used in the Middle Ages. Its
precursors include e-cash and digicash invented by D. Chaum, and bitgold
invented by N. Szabo, see \cite{Chaum}, \cite{Popper}.\footnote{%
There is a heated debate of the true idenity of Satoshi Nakamoto. Nick Szabo
is often mentioned as a potential inventor of bitcoin. Here is a small piece
of evidence, which might be of interest. Nakamoto's initials are SN, while
Szabo's are NS. However, Szabo is originally a Hungarian name, where the
last name comes first, so his initials would be SN. An interesting
coincidence.}

All building blocks of bitcoin ecosystem have been known for some time,
including two of the most important techniques in public-key cryptography,
the Diffie-Hellman key exchange protocol and the RSA encryption system, see 
\cite{Ellis}, \cite{Diffie}, \cite{Rivest}.\footnote{%
While these techniques were discovered in the academic community in 1976 and
1978 respectively, they were known in the intelligence community since at
least 1974.} Proof of work, based on cryptographic hash functions,
specifically SHA-256, is similar to hashcash invented by Back, \cite{Back};
while Merkle trees were introduced in the seminal paper by Merkle, \cite%
{Merkle}.

Ignoring such nuances as wallets, etc., we can describe the basic setup as
follows. Participants of the system are represented by their public/private
key pairs. The main control variable is the number of bitcoins belonging to
a particular public key. This number is known to all participants at all
times (in theory). The owner of a particular public key broadcasts their
intend to send a certain quantity of bitcoins to another public key. Miners
aggregate individual transactions into blocks, verify them to ensure that
there is no double spend by competitively providing proof of work, and
receive mining rewards in bitcoins. A\ transaction is confirmed if there are
at least six new blocks built on the top on the block to which it belongs. A
typical block is shown in Figure \ref{Fig 3}.

The size of mining rewards is halved at regular intervals; so that the total
number of bitcoins in circulation converges to 21 Mil. Currently there are
about 16 Mil bitcoins in circulation. It is believed that 3-5 Mil are
irretrievably lost. Some 450,000 blocks have been mined so far; a new block
is mined every ten minutes on average. Due to the fact that mining rewards
are paid with \emph{new} bitcoins, transaction costs are claimed to be very
low. This is a nifty bit of sleight of hand however, because the value of
existing bitcoins is constantly diluted. Some representative bitcoin
statistics is given in Figure \ref{Fig 4}.

Bitcoin promises are grand. Its proponents expect it to become a
supra-national currency eventually supplanting national currencies, which,
in their minds, can be easily manipulated. Many even believe that bitcoin is
the modern digital version of gold, due to the effort required for PoW, see,
e.g., \cite{Popper}. Whilst bitcoin is clearly an impressive breakthrough,
reality is much less grand than perception, and is quite telling:

(A) A new block is created on average every 10 min.;

(B) The number of transactions per second (TpS) is approximately 7, compared
to 2,000 TpS on average handled by VISA;

(C) In monetary terms, the amount of transactions is about 100 Mil USD/day;

(D) Current real (not nominal!) transaction costs are 1.5 Mil USD/day, 1.5\%
of total volume; in 2012 it was whopping 8\%, in 2014 - 6\%;

(E) Mining is a cost of electricity game. In high energy cost countries
miners go bust: Swedish KnCMiner recently declared bankruptcy ahead of
halving miner's reward. While exact numbers are not known, it is believed
that bitcoin consumes as much electricity as EBay, Facebook and Google
combined;

(F) Miners are arranged in gigantic pools (so much for P2P mining!); AntPool
- 18.7\%, F2Pool - 17.7\%, BitFury - 7.7\%, BTCC Pool - 7.4\%, BW.COM -
7.3\%. Thus, a 51\% attack becomes possible! There is a very \emph{high}
probability that six consecutive blocks will be mined by the same actor (so
much for checks and balances!). Most of all these pools are Chinese, partly
due to low electricity cost, partly due to high tech advances. Not only
miners are predominantly Chinese, so are the players - 91\% CNY, 7\% USD,
1\% EUR;

(G) At the moment, the main purpose of using bitcoin is for speculation and
circumvention of capital controls in China.

It is truly amazing to see how miners are prepared to perform socially
useless tasks, as long as they are paid for it. A telling historical analogy
jumps to mind: During the contest for design of the dome of Santa Maria del
Fiore, it was suggested to use dirt mixed with small coins to serve as
scaffolding. After the dome's completion\ the dirt was to be cleared away
for free by the profit-seeking citizens of Florence (proto-miners). It is
clear that BC/DL is still awaiting its Brunelleschi, \cite{King}.

T. J. Dunning, quoted by Karl Marx in \textit{Das Kapital}, \cite{Marx}, put
it succinctly:

\begin{quote}
"With adequate profit, capital is very bold. A certain 10 per cent. will
ensure its employment anywhere; 20 per cent. certain will produce eagerness;
50 per cent., positive audacity; ..."
\end{quote}

\section{Potential usages of DLT in banking}

\label{DLTinbanking}

\subsection{Banking X-Road}

No bank, however big, is an island; banks can only operate as a group. In
the process of their day-to-day activities, they become naturally
interlinked. Due to these linkages between banks, DLT can provide a useful
tool for facilitating, reconciling, and reporting their interactions. Given
that internal technology is bank specific, it is impractical to standardize
bank infrastructure. However, it is possible to bring them to a common
denominator by emulating the success of the Estonian X-Road and creating a
DL solution for banking operations, which, by analogy, can be called the
e-bank X-Road. In this regard, DL will serve as an adapter, not dissimilar
to an electrical adapter.

In 1997 Estonia started to move to digital government. In 2001, A. Ansper in
his master thesis, \cite{Ansper}, proposed a suitable design. He developed a
distributed P2P secure information system called the e-Estonia X-Road based
on the idea of adaptor. X-Road is the digital environment which links
various heterogeneous public and private databases and enables them to
operate in sync. A small company Cybernetica implemented this design for
around 60 Mil EUR.\footnote{%
Other countries tried to follow suite but not all attempts were unqualified
success.}

Let's describe a possible design for the e-bank X-Road. Given the
non-scalable nature of PoW, and unclear security properties of PoS, X-road
has to be controlled by trusted notaries or validators. Two financial
institutions, represented by their public keys, use their respective
adapters to agree on common terms on a deal. They digitally sign and execute
a smart contract, hash it, and broadcast the hashed version to the X-Road
participants. A quorum of notaries digitally signs the hash ("laminates"
it), and re-posts the signed hash in the common X-Road layer. Validators are
paid for their services, similarly to central securities depositories.%
\footnote{%
Corda, recently described in a white paper by R3, might be a step in this
direction, \cite{Brown}.}

It is worth noting that a blockchain does not by itself guarantee
unambiguous ownership: steps are required to identify and resolve any
ambiguities before moving to a BC, and in addition, tools and mechanisms to
resolve ambiguities only discovered when the BC is already well established.
Both of these requirements are underemphasized in current discussions of
BC/DLT applications.

There are several smaller areas in which DLT can be used to reduce
transaction costs and other frictions in the conventional system. Such areas
include but are not limited to:

(A) post-trade processing;

(B) global payments;

(C) trade finance;

(D) rehypothecation;

(E) syndicated loans;

(F) real estate transactions.

\subsection{Trade execution, clearing, settlement}

The holy trinity of capital markets is trade execution, clearing,
settlement. While initial public offering of stock is an important rite of
passage for a new company, secondary trading is a mechanism for continually
re-allocating ownership and control in a somewhat optimal fashion. In
addition to stocks, many other products, such as equity derivatives,
interest rate swaps, commodities, etc. are traded on public exchanges.
Moving many over-the-counter (OTC) products to exchanges is an important
regulatory imperative \cite{Skeel}.

Currently, there are three necessary steps to trade public securities:

(A) Buyers and sellers have to be matched;

(B) The transaction has to be cleared, i.e. novated to a central clearing
counterparty (CCP);

(C) The transaction has to be settled, i.e. delivery vs. payment (DvP) has
to take place; so that title and money can be transferred as expected.

These steps are characterized by vastly different time scales - trading
often takes place in milliseconds, while clearing and settlement take 1-3
days! Although the proverbial T+2, T+3 irritate many people, they might be a
bit too fast to push for the T+15' solution. The actual process is very
involved and includes investors, custodial banks, exchanges, brokers
(general clearing members of CCPs), CCPs, central securities depositories,
regulators, etc.

It is natural to ask if a different design of exchanges can improve the
overall process and make it more stable and less costly. The answer is yes
and no. On the pros side, there are several issues which the current set-up
solves very well:

(A) counterparty credit risk management;

(B) netting;

(C) DvP and credit risk more generally, which is addressed by collecting
Initial Margin, Variation Margin, and Guarantee Fund contribution from
clearing members;

(D) anonymity;

(E) ability to borrow stocks.\footnote{%
The thriller \textquotedblleft Ronin\textquotedblright\ was not universally
critically acclaimed: some critics struggled to identify what it was about 
\cite{Turan}. In the author's view, it takes the difficult challenges of
transactions amoung many untrustworth parties which underlie many great
thrillers and brings them to the fore, making \textquotedblleft
Ronin\textquotedblright\ arguably the greatest of all thrillers (perhaps the
ending would have been different had the characters know about blockchain).}

On the cons side, numerous issues are rather disconcerting:

(A)\ cost;

(B) speed;

(C) need for reconciliation and failures.

It is clear that straightforward attempts to apply a blockchain to clearing
and settlement (thankfully, to the best of the author's knowledge, nobody
wants to use it in trading \textit{per se}) cannot be successful. The
reasons are simple - instantaneous settlement (T+15' as it is occasionally
called)\ obliterates all the aforementioned advantages of the current
system. It increases the money sloshing around by at least an order of
magnitude. Thus, slow clearing and settlement is not so much a consequence
of the technological backwardness of exchanges and CCPs (although they are
not always using cutting edge technology), but rather a result of their 
\textit{modus operandi}.

By using permissioned private ledger(s) one can certainly cut costs,
somewhat increase speed of clearing and settlement, and reduce the number of
failures and hence the need for reconciliation. In particular, smart
contracts, if they can be legally enforced, can solve a \emph{part} of the
DvP conundrum, which will require that \emph{both securities and cash} are
parts of the same ledger. While smart contracts cannot solve all problems,
they represent a step in the right direction. A potential evolution of the
trading-clearing-settlement triad is illustrated in Figure \ref{Fig 5}.

\subsection{Global payments, trade finance, rehypothecation}

Global payments is another area, where DLT can be potentially useful. It is
important to note that, in spite of claims to the contrary, the payment
system \emph{is not broken }but rather expensive. For instance, Real-Time
Gross Settlement system works well for domestic transactions but is
inefficient and expensive for foreign transactions. Thus, some synergies can
be gained if a DL, which supports several national currencies at once, is
developed to replace the legacy system.

For trade finance, there is the potential to use BC/DL to simplify the flow
of information among all participants and smart contracts to partially solve
the DvP problem.

In the rehypothecation set-up, it is possible to use BC/DL to untangle the
ownership of the collateral. However, this is more of an accounting tool,
rather than a comprehensive solution because in many instances the actual
legal ownership of collateral cannot be established with certainty.

\section{Monetary circuit and money creation}

\label{Circuit}

\subsection{Monetary circuit}

For centuries, the origins, properties and functions of money have been
debated in countless expositions. In the fourteenth century, the sagacious
French abbot Gilles li Muisis lamented, \cite{Bloch}:

\begin{quote}
\textquotedblleft Money and currency are very strange things; They keep on
going up and down and no one knows why; If you want to win, you lose,
however hard you try.\textquotedblright\ 
\end{quote}

In the twentieth century the great British economist John Maynard Keynes
shrewdly observed, \cite{Keynes}:

\begin{quote}
\textquotedblleft For the importance of money essentially flows from it
being a link between the present and the future.\textquotedblright\ 
\end{quote}

As was mentioned earlier, money is inherently linked with banking, which,
over many centuries, gradually evolved from full-reserve towards fractional
reserve banking. For instance, the Bank of England founded in 1694 already
operated as a fractional reserve bank.\footnote{%
The Bank of England was characterized by Marx, \cite{Marx}, as follows:
\par
\begin{quote}
"At their birth the great banks, decorated with national titles, were only
associations of private speculators, who placed themselves by the side of
governments, and, thanks to the privileges they received, were in a position
to advance money to the State. Hence the accumulation of the national debt
has no more infallible measure than the successive rise in the stock of
these banks, whose full development dates from the founding of the Bank of
England in 1694."
\end{quote}
}

In modern societies commercial banks are almost exclusively fractional and
produce money "out of thin air", see, \cite{Keen}, \cite{Werner}, \cite%
{Lipton}. This important fact is thoroughly misunderstood by the modern
macroeconomic thinking, which incorrectly overemphasizes the intermediation
aspect of banking and assigns the money creation role to central banks
instead of commercial banks. In reality, commercial banks are not
constrained by their deposits and can and do issue money at will. At the
same time, their ability to do so is restricted by banking regulations,
which impose floors on the amount of banks' capital and liquidity, so that
money creation cannot go \textit{ad infinitum}.

To understand the role played by money in the economy, one needs to follow
its flow and to account for non-financial and financial stocks (cumulative
amounts), and flows (changes in these amounts). Here is how Michal Kalecki,
the great Polish economist, summarizes the complexity of the issues at hand
with his usual flair and penchant for hyperbole, \cite{Robinson}:

\begin{quote}
\textquotedblleft Economics is the science of confusing stocks with
flows.\textquotedblright\ 
\end{quote}

In the author's opinion, the functioning of the economy and the role of
money is best described by monetary circuit theory (MCT), which provides a
unifying framework for specifying how money lubricates and facilitates
production and consumption cycles in the society. MCT describes in the most
precise way the dynamics of the economy and explains how and by whom money
is created. More specifically, it describes the interactions among the five
sectors, including government, central bank, private banks, firms, and
households. As part of the MC, private banks play an outstanding role as
credit money creators. In this framework, central banks don't create money
directly, but rather accelerate or slow down the process of money creation
by private banks, by providing a unique universal medium in the form of
electronic cash for different banks to control their inventories of assets
and liabilities. A\ schematic representation of the monetary circuit is
given in Figure \ref{Fig 6}, which represents money flowing among the above
mentioned five sectors of the economy.

\subsection{General aspects of money creation}

Currently, there are three theories explaining money creation: the credit
creation theory, the fractional reserve theory, and the financial theory of
intermediation, see, e.g., \cite{Keen}, \cite{Werner}, \cite{Lipton} and
references therein. The author firmly believes that only the credit theory
advocated by Macleod, Hahn, Wicksell, and Keen among others, correctly
reflects the mechanics of linking credit and money creation. Credit creation
theory was popular in the nineteenth century, but, unfortunately, gradually
lost ground and was overtaken by the fractional reserve theory of banking,
which, in turn, was supplanted by the financial theory of intermediation. In
the author's view, the latter theory severely underemphasizes the unique and
special role of the banking sector in the process of money creation, and
cannot rationally explain things like the global financial crisis of
2007-2008 and other similar events, which happen with disconcerting
regularity. This aspect is particularly important because currently there is
a profound lack of appreciation on the part of the conventional economic
paradigm of the special role of banks. For example, banks are excluded from
widely used dynamic stochastic general equilibrium models, which are
influential in contemporary macroeconomics and popular among central
bankers, in spite of the fact that they systematically fail to produce any
meaningful results \cite{Buiter}. It is clear that a vibrant financial
system cannot operate without banks, and that the banking system is very
complex and difficult to regulate because banks become interconnected as a
part of their regular lending activities. In addition to their money
creation role, banks regulate access to the monetary system, by providing
KYC and AML services.

\subsection{Money creation by individual banks}

We start with the simplest situation, and consider a single bank, which
lends money to a borrower who immediately deposits it with the same bank.
Thus, the bank simultaneously creates assets and liabilities. The size of
the loan is limited solely by regulations and bank's own risk appetite. The
full cycle from money creation to money annihilation is shown in Figure \ref%
{Fig 7}. Money is pumped into the system (created) when it is lent out by
the bank and pumped out (annihilated) when it is repaid. If the borrower
repays, the principal is destroyed, but the interest stays in the system. If
the borrower defaults, the money stays in the system indefinitely. The chain
of money transfers from one owner to the next is naturally described by a
BC, ideally residing on DL.

\subsection{Money creation by the banking system}

A\ more complex case of asset creation by one bank and liabilities by a
second bank is illustrated in Figures \ref{Fig 8}, \ref{Fig 9}. Linkages
between these two banks occur because the first one has to borrow cash from
the second, so that their central bank cash holdings reach suitable levels.
In this setup, it is clear that central banks do not generate money
themselves; instead, they play the role of liquidity providers (if, for
example, the second bank does not want to lend money to the first), and
system stabilizers (similar to\ the Watt's centrifugal governor). Thus,
central banks are the glue, which keeps the financial system together. It is
clear that BC is even more relevant in the case in question.

\subsection{Bank lending vs. bitcoin and P2P lending}

In view of the above, the key distinction between bank money\ creation and
bitcoin mining, P2P lending, etc., is evident. Banks create money "out of
thin air". Since bitcoin transactions are not based on credit, they simply
move existing money around. Same is true for P2P transactions - P2P
operators are strictly intermediaries, they don't create money at all!
Therefore, banks and P2P operators lend on different scales: banks -- money
they don't have, P2P - only money they have. Hence, the P2P impact on the
financial system as a whole is very limited.

\section{CBDC and negative interest rates}

\label{CBDC}

\subsection{Why CBDC?}

Can and should central banks issue DC? Recently, these discussions have been
invigorated by the introduction of bitcoin, \cite{Nakamoto}, and a
persistence of negative interest rates, which plagued Medieval Europe in the
form of demurrage,the Brakteaten system, and numerous variations of the same
tune for centuries. Recall that demurrage was a tax on monetary wealth and
required a massive apparatus of coercion to be imposed efficiently. Today,
even the best-in-class economists seem to be unsure of its true nature; for
instance, Prof. Rogoff equates it with currency debasement, which is a very
different mechanism, \cite{Rogoff}. The idea of scrip money, i.e. money
which requires paying of periodic tax to stay in circulation, thus emulating
demurrage, was proposed by S. Gesell, the German-Argentinian entrepreneur
and self-taught economist, in the febrile post-WWI atmosphere, \cite{Ilgmann}%
. Subsequently, it was regurgitated by Irving Fisher during the great
depression, \cite{Fisher}.

In the author's view, it is a sad reflection of the current state of
economic affairs, and the level of economic insight, that the current low
interest rate environment has prevailed for such a long time, in spite it
being such an ineffective tool. Moreover, in some economies, such as
Switzerland and Denmark, interest rates have reached seriously negative
levels.\footnote{%
One cannot help but notice with a modicum of satisfaction, that critics of
the celebrated Vasicek model for interest rates, \cite{Vasicek}, who
vigorously attacked him for allowing short rates to become negative, proved
to be completely wrong.}

Negative interest rates can be used to simulate inflation; the crucial
difference between these two regimes is that physical cash is very valuable
under the former, and highly undesirable under the latter. The last line of
defense between us and meaningfully negative rates is paper currency.
However, in many societies, particularly in Scandinavia, cash is relegated
to the far corners of the economy already. It is not hard to imagine that in
a few years' time instead of banknotes, we shall have CBDC, \cite{Barrdear}, 
\cite{Broadbent}, \cite{Lipton2}. Once cash is abolished, interest can be
made as negative as desired by central bankers.

\subsection{How CBDC can be issued?}

Currently, there are two approaches to creating digital currencies on a
large scale. The first one, which has gained popularity since the invention
of bitcoin, is based on unpermissioned DL, whose integrity is maintained by
notaries (or miners), see, e.g., \cite{Danezis}. Participants in this BC are
pseudo-anonymous since they are hidden behind their public keys. However, in
principle, they can be identified by various inversion techniques applied to
old recorded transactions, \cite{Reid}.

An earlier approach was developed by Chaum, who introduced a blind signature
procedure for converting bank deposits into anonymous cash, see \cite{Chaum}%
. Chaum's approach is much cheaper, faster and more efficient than the
bitcoin-style one. However, it heavily relies on the integrity of the
cash-issuing bank rather than on trustless integrity of bitcoin secured by
computational efforts of miners. Central banks can follow either avenue for
issuing digital cash. By doing so, central banks will be indirectly
providing access to their balance sheets to general public. However, in
either eventuality, central banks won't be able to perform KYC and AML
functions and would still have to rely on commercial banks, directly or
indirectly, for doing so.

One possibility is as follows: a central bank issues numbered currency units
into DL, whose trust is maintained by designated notaries receiving payments
for their services. Thus, at any moment, there is an immutable record
showing which public key is the owner of a specific currency unit. Given
that notary efforts are significantly cheaper and faster than that of
bitcoin miners, this construct is easily scalable to satisfy needs of the
whole economy. Moreover, since the records of transactions are immutable, it
is possible to de-anonymize transactions thus maintaining AML requirements.

In summary, modern technology makes it possible to abolish paper currency
and introduce CBDC, which can also be used to address some of the societal
ills, such as crime, drug trafficking, illegal immigration, etc., and
eliminate costs of handling physical cash, which are of order of 1\% of the
country's GDP. It will smooth the motion of the wheels of commerce and help
the unbanked to become participants in the digital economy, thus positively
affecting the society at large.

\subsection{How CBDC\ can be used to implement the Chicago Plan}

Moreover, CBDC makes the execution of the celebrated Chicago Plan of 1933,
originally proposed by D. Ricardo in 1824, for introducing narrow
(full-reserve) banking entirely possible - both firms and ordinary citizens
can have accounts directly with central banks, thus negating the need of
having deposits with commercial banks, see \cite{Allen}, \cite{Benes}, \cite%
{Baynham}, \cite{King1}. In this case, banks will lose their central
position in the economy and become akin to utility providers. They would
have to maintain the amount of central bank cash equal to the amount of time
deposits. Such narrow banks would in essence become the guardians of the
system by providing KYC and AML services and executing simple transactions.
In fact, in the wake of the global financial crisis, many central banks
massively increased their balance sheets, while commercial banks have chosen
to keep enormous quantities of non-mandatory deposits with them. Thus, the
system \textit{de facto} has moved towards narrow banking.

\section{Conclusions}

\label{Conclusions}While the idea of BC/DLs is not new, modern technology
gives it a new lease of life. DLT opens new possibilities for making
conventional banking and trading activities less expensive and more
efficient by removing unnecessary frictions. Moreover, if built with skill,
knowledge, and ambition, it has potential for restructuring the whole
financial system on new principles. We emphasize that achieving this goal
requires overcoming not only technical, but also political obstacles.

While DLT has numerous applications, it is not entirely clear, which
financial applications should be handled first. Exchanges, payments, trade
finance, rehypothecation, syndicated loans, and other similar areas, where
frictions are particularly high, are attractive candidates. DC, including
CBDC, is another very promising venue.

Currently, many applications of DL and related technology appear to be
misguided. In some cases, they are driven by a desire to apply these tools
for their own sake, rather than because the result would be clearly
superior. In other cases they are driven by a failure to appreciate that the
current systems may not be as they are because of technological reasons, but
rather because of business and other reasons.

So far, practical application of DLT in finance have been limited and a lot
remains to be done in order to achieve real breakthroughs.

\begin{acknowledgement}
The invaluable help of Marsha Lipton from Numeraire Financial in thinking
about and preparing this presentation cannot be overestimated. I am grateful
to several colleagues, including Alex Pentland and David Shrier from MIT,
Damir Filipovic from EPFL, Matheus Grasselli from McMaster, and Paolo Tasca
from UCL for their help and suggestions. As a CEO of StrongHold Bank Labs, I
am currently working on a new type of a digital bank, which will be
utilizing some of the ideas presented in this paper. I am grateful to my
colleague Julian Phillips from\ SHBLabs for his insightful and thorough
feedback.
\end{acknowledgement}

\begin{figure}[tbph]
{\ \center%
\includegraphics[width=1.0\textwidth, angle=0]
{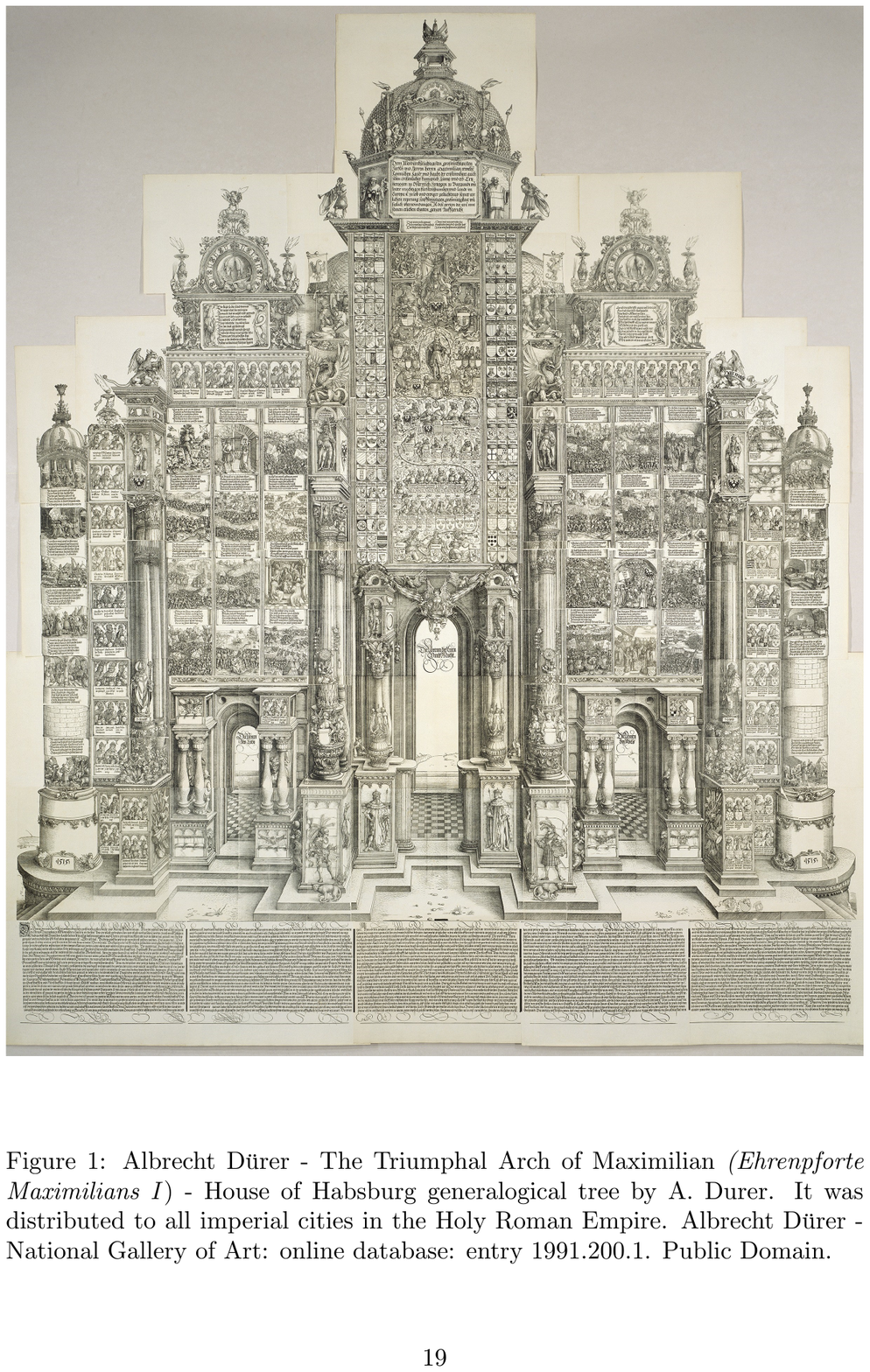} }
\label{Fig 1}
\end{figure}

\begin{figure}[tbph]
{\ \center%
\includegraphics[width=1.0\textwidth, angle=0]
{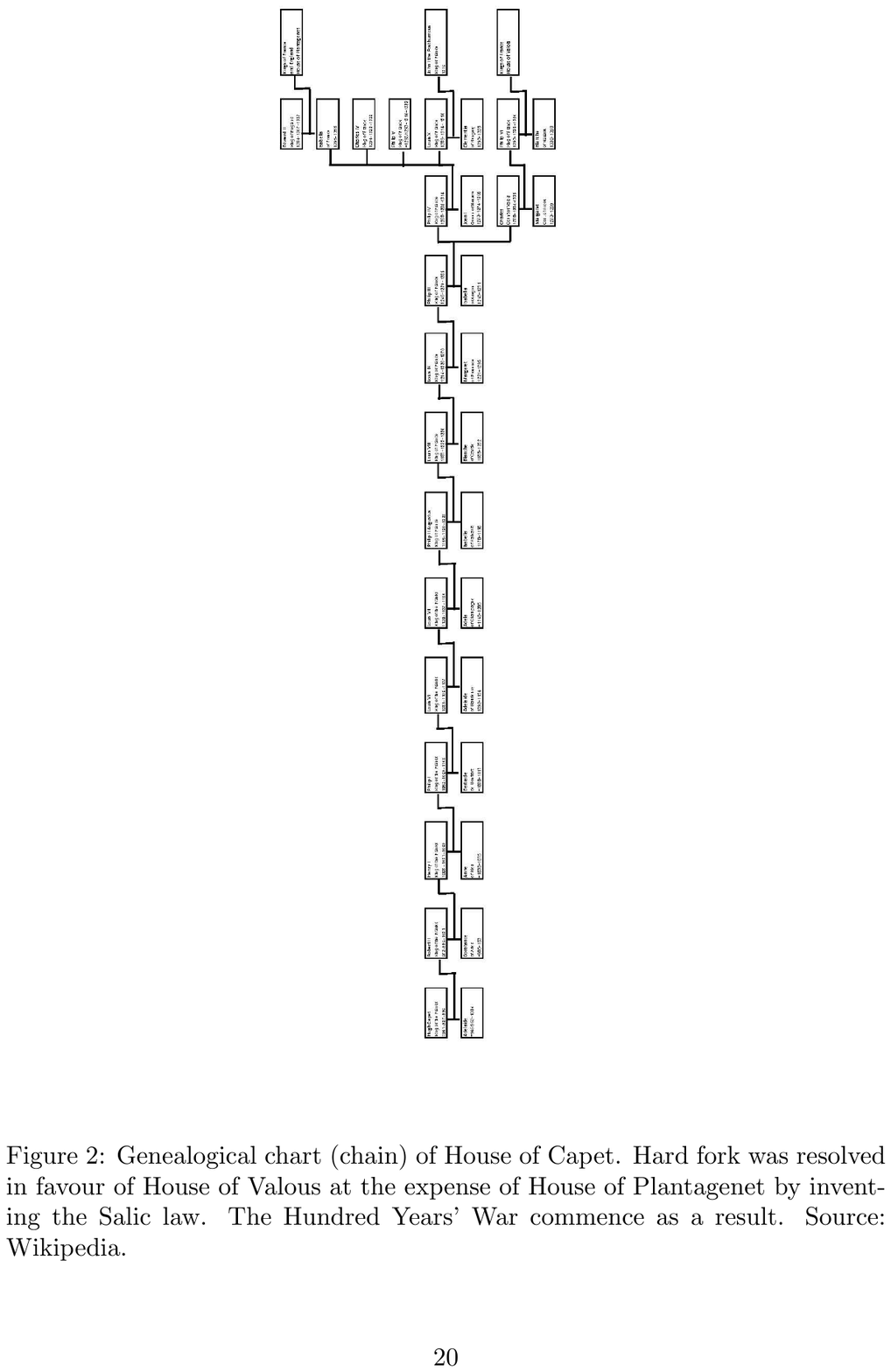} }
\label{Fig 2}
\end{figure}

\begin{figure}[tbph]
{\ \center%
\includegraphics[width=1.0\textwidth, angle=0]
{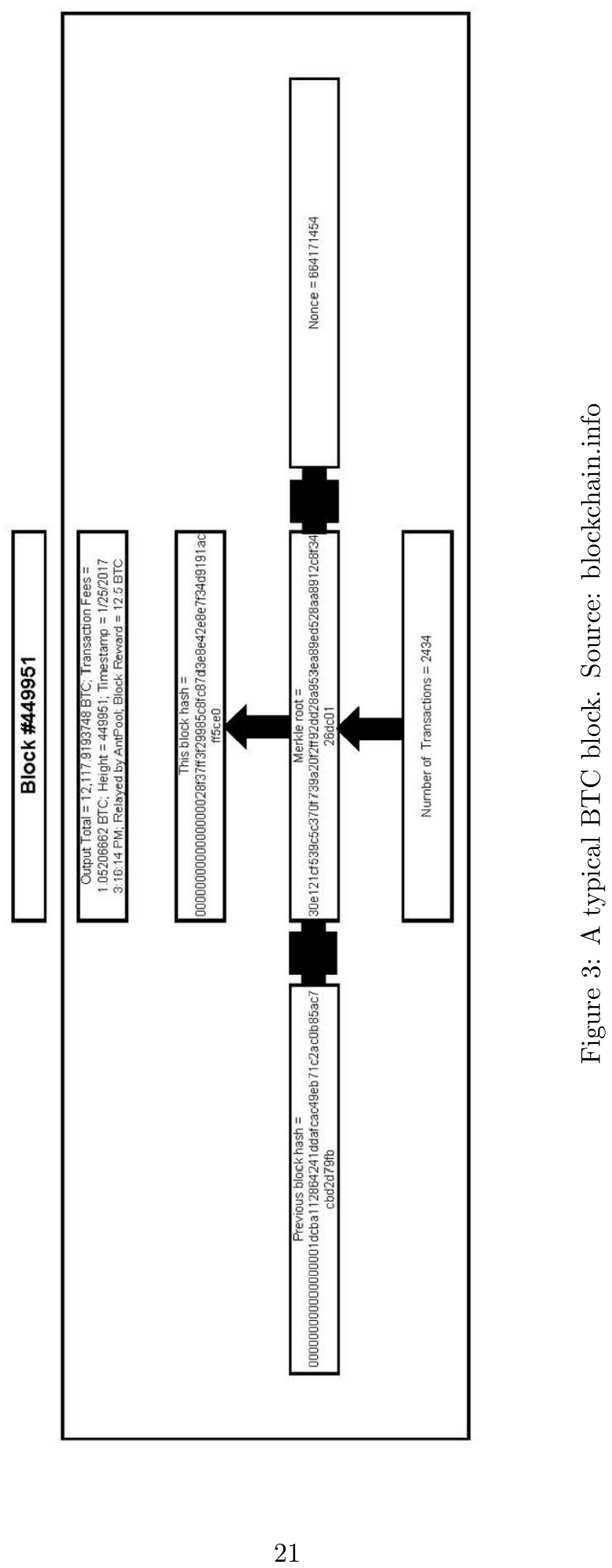} }
\label{Fig 3}
\end{figure}

\begin{figure}[tbph]
{\ \center%
\includegraphics[width=1.0\textwidth, angle=0]
{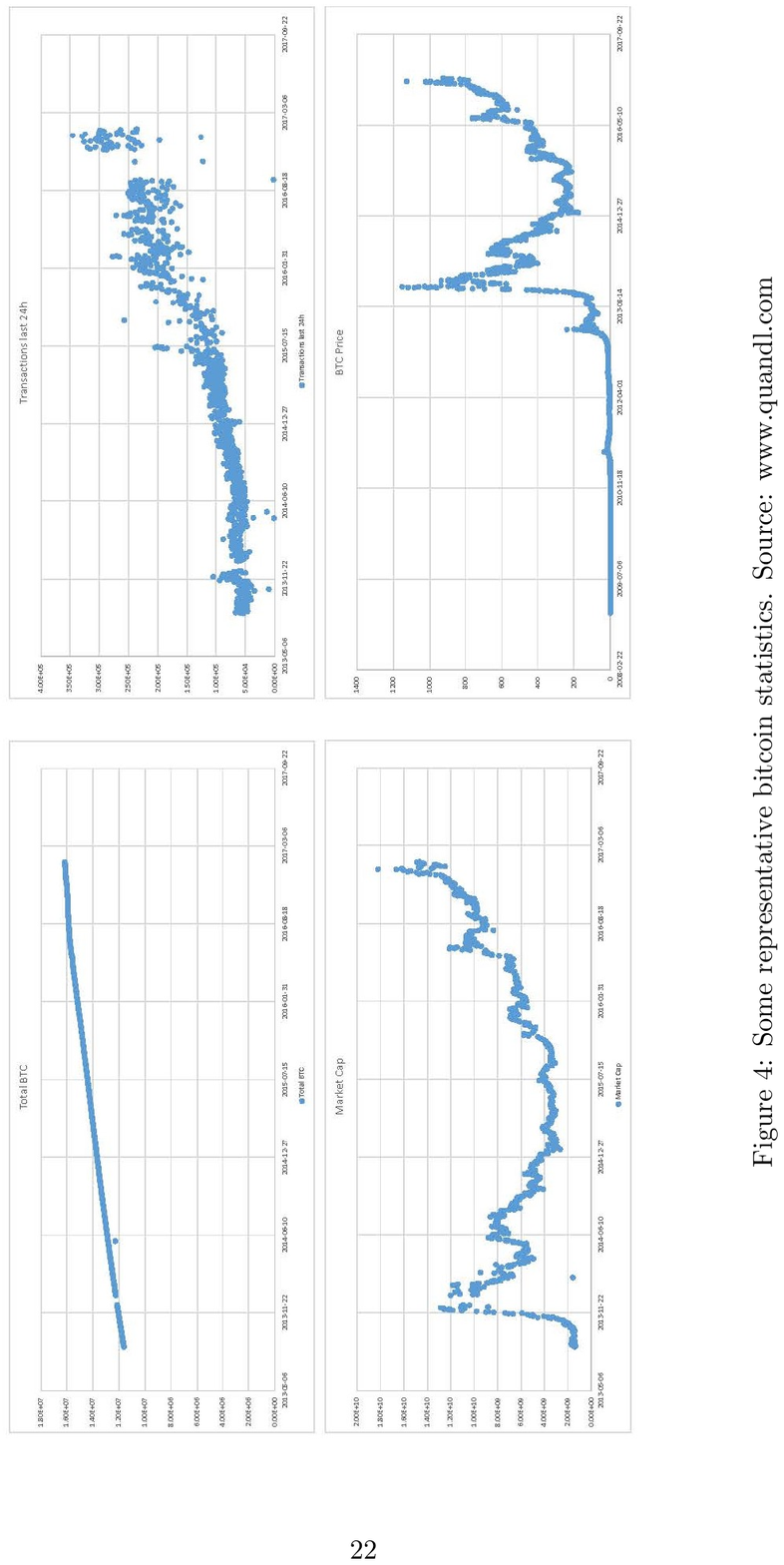} }
\label{Fig 4}
\end{figure}

\begin{figure}[tbph] 
{\ \center%
\includegraphics[width=1.0\textwidth, angle=0]
{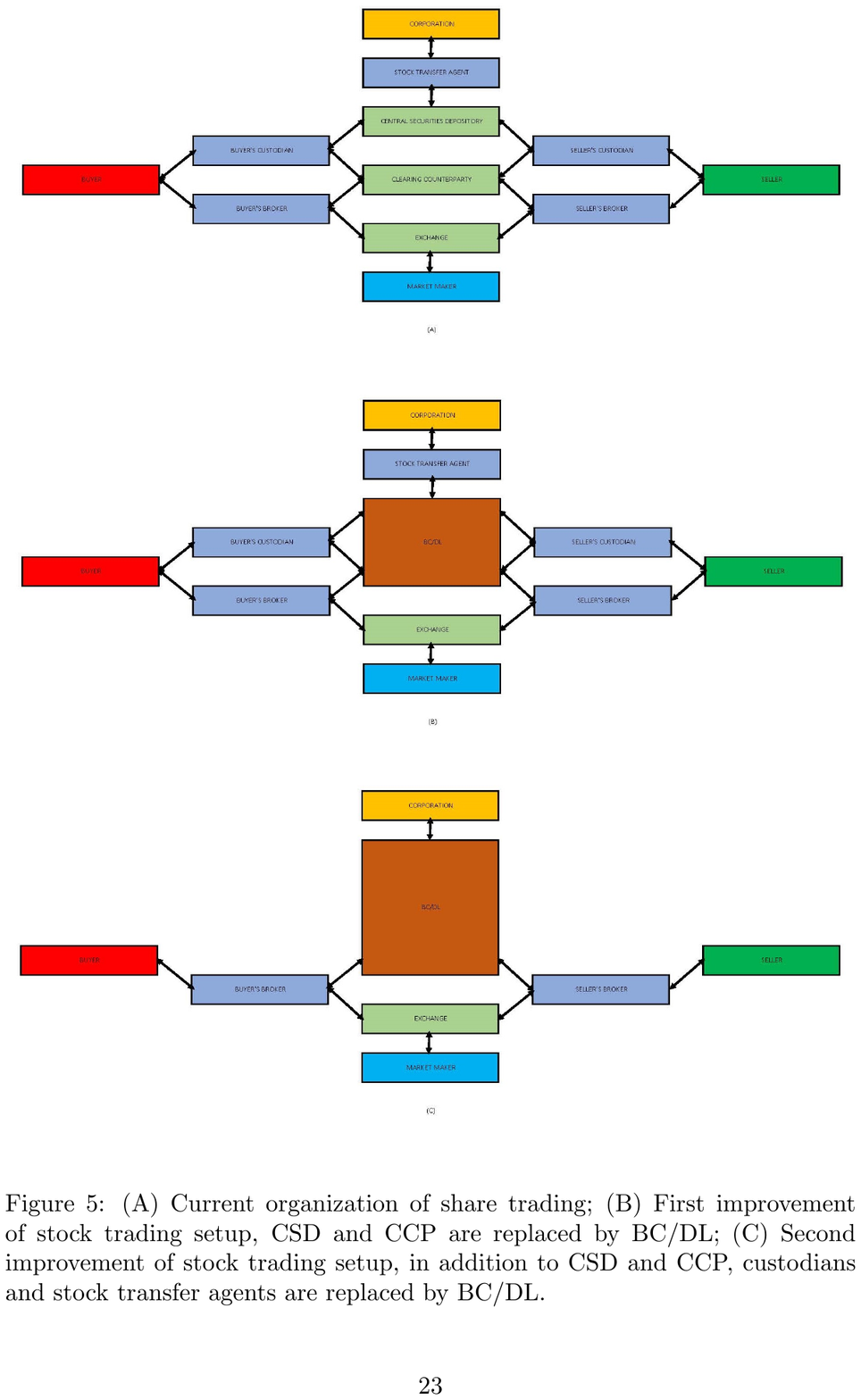}}
\label{Fig 5}
\end{figure}

\begin{figure}[tbph]
{\ \center%
\includegraphics[width=1.0\textwidth, angle=0]
{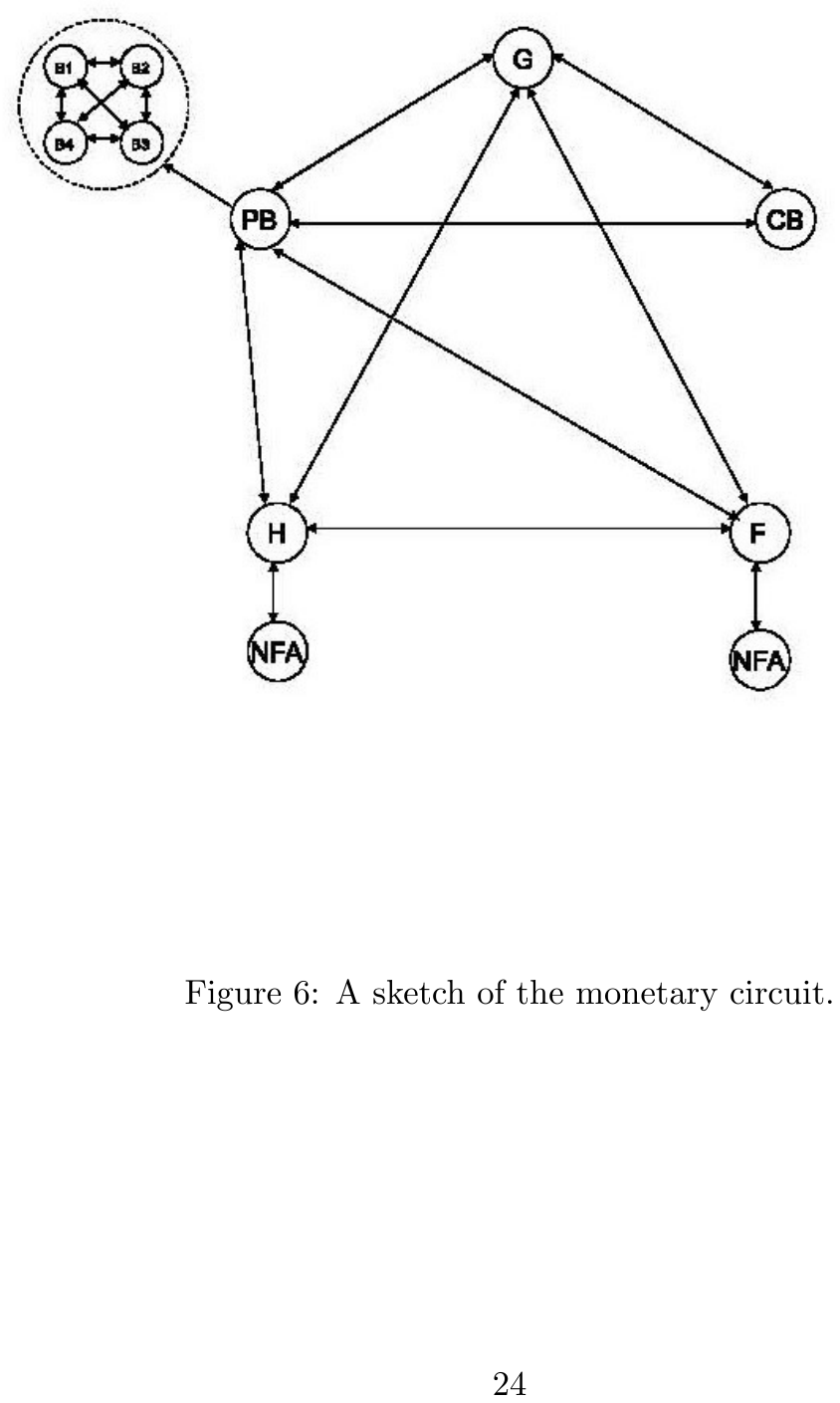} }
\caption{A sketch of the monetary circuit.}
\label{Fig 6}
\end{figure}

\begin{figure}[tbph]
{\ \center%
\includegraphics[width=1.0\textwidth, angle=0]
{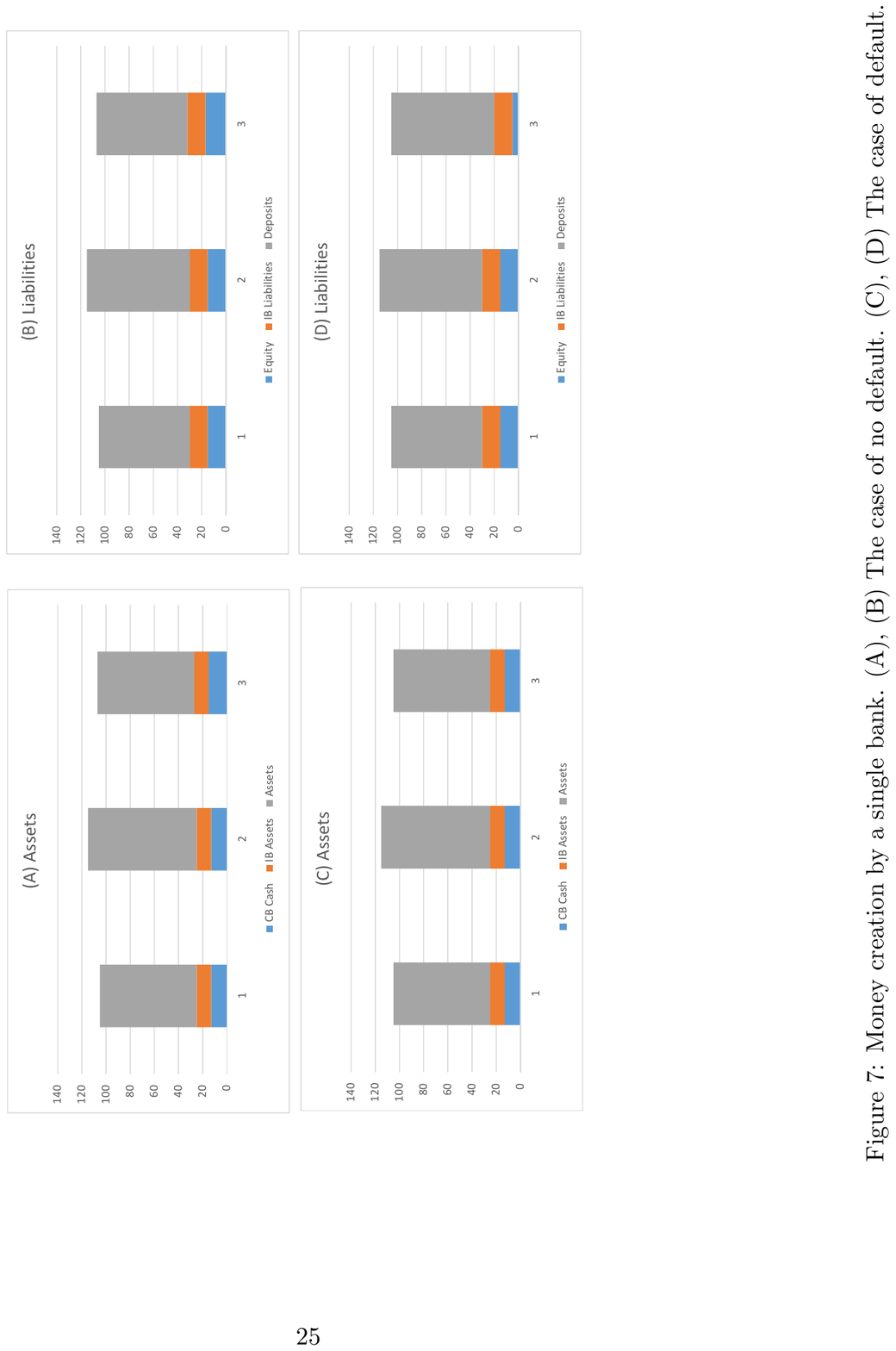} }
\label{Fig 7}
\end{figure}

\begin{figure}[tbph]
{\ \center%
\includegraphics[width=1.0\textwidth, angle=0]
{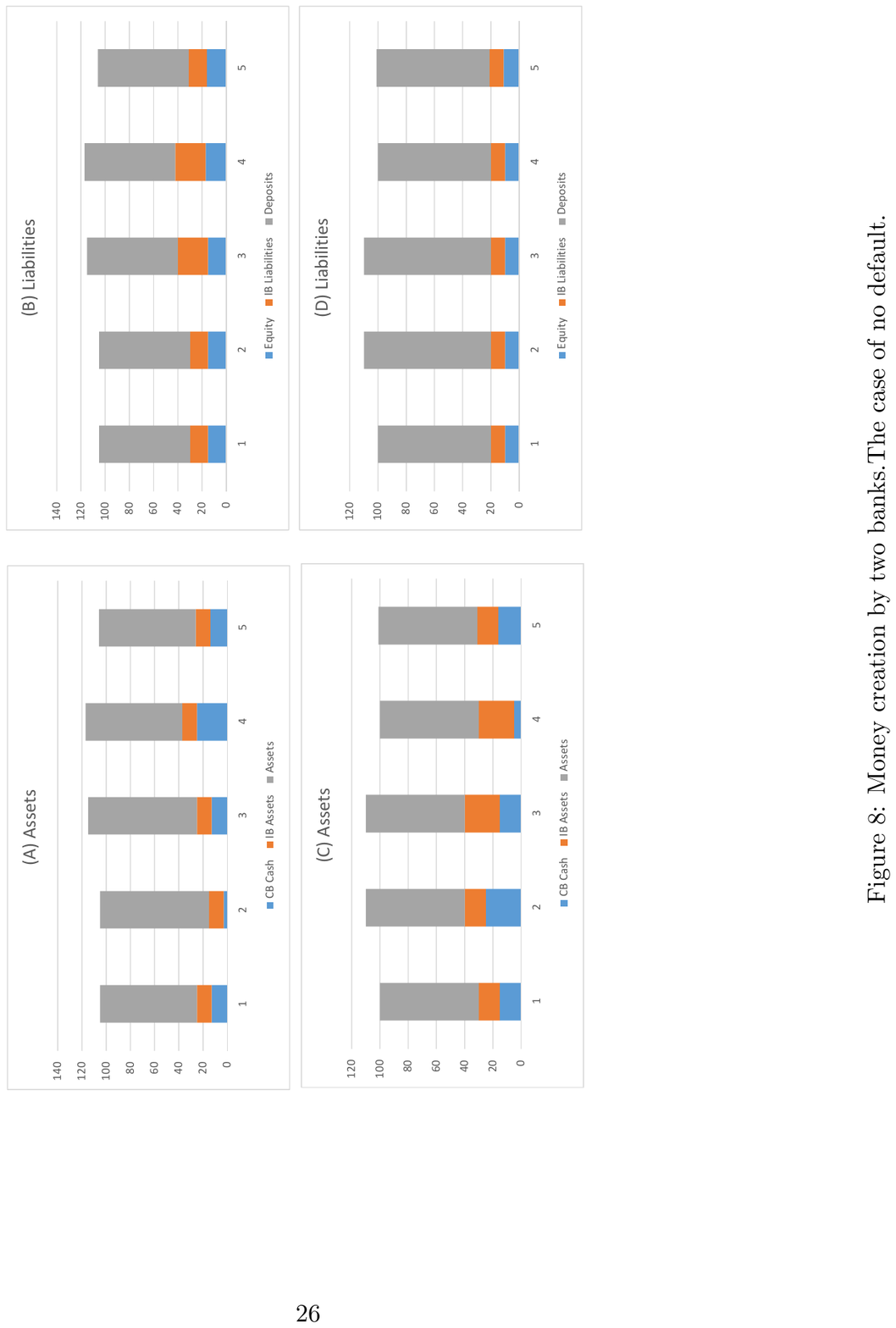} }
\label{Fig 8}
\end{figure}

\begin{figure}[tbph]
{\ \center%
\includegraphics[width=1.0\textwidth, angle=0]
{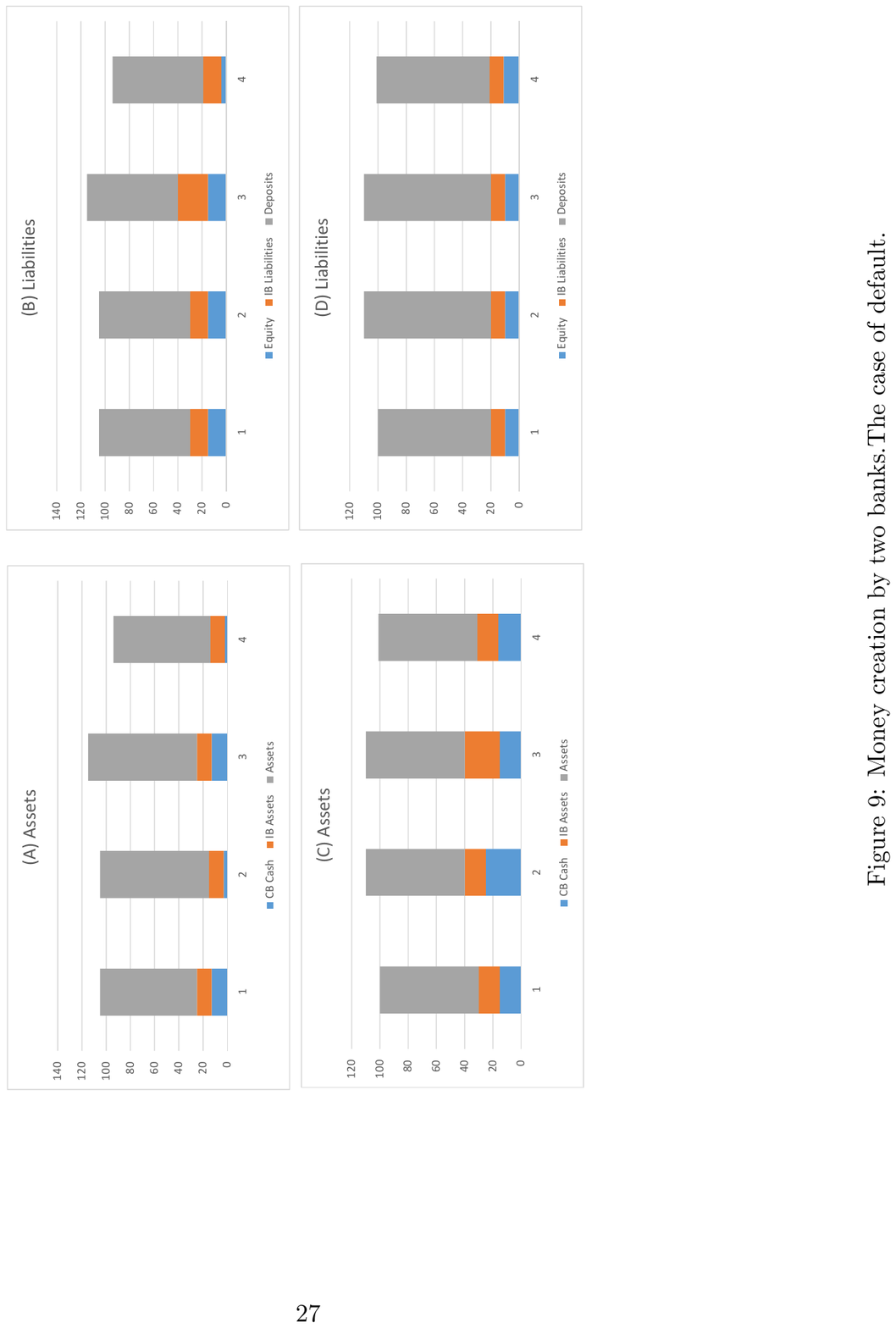} }
\label{Fig 9}
\end{figure}

\end{document}